\documentclass[default,times]{aastex62}

\usepackage{color}
\usepackage{hyperref}
\hypersetup{colorlinks=true, linkcolor=blue}

\graphicspath{{./}{figures/}}

\shorttitle{Lessons Learned from Virtual Transitions}
\shortauthors{Candelaria, Hunt, Rowland}

\begin{document}

\title{Lessons Learned from Virtual Transitions of Undergraduate Student Support Programs due to the COVID-19 Pandemic}


\author[0000-0002-7517-5337]{Tierra Candelaria}
\affiliation{New Mexico Institute of Mining and Technology, 801 Leroy Pl, Socorro, NM 87801, USA}

\author[0000-0002-4669-0209]{Qiana Hunt}
\affiliation{University of Michigan, 1085 S University, Ann Arbor, MI 48109, USA}

\author[0000-0003-2144-4885]{Danielle Rowland}
\affiliation{Department of Astrophysics, American Museum of Natural History, 200 Central Park West, New York, NY 10024, USA}

\begin{abstract}
The COVID-19 global pandemic spurred sudden and dramatic changes in the way universities and research programs operate, often with negative consequences that have disproportionately affected minorities, women, and people with low income. These consequences compound the difficulties these groups faced prior to the pandemic | in particular, the increased obstacles marginalized students encounter while pursuing STEM career paths. The National Astronomy Consortium (NAC) was founded to support underrepresented physics and astronomy students by providing research opportunities and long-term mentoring. In response to the pandemic and nation-wide shutdowns, the NAC shifted both its summer research program and its annual fall conference to a fully-remote format. Here, we discuss the changes the NAC made to achieve a successful virtual transition, and provide temporary and long-term recommendations for programs seeking to provide support to marginalized undergraduate and graduate researchers moving forward.  
\end{abstract}


\section{Introduction}
\label{section:introduction}

Even before the novel coronavirus escalated into the world-wide pandemic it is today, institutions have struggled to get a handle on the widespread disparities that persist in academia. 
There is much debate over how to recruit and retain junior academics who are historically excluded from their scientific communities along lines of race, gender, and economic status. 
Summer research experiences for undergraduates have become a crucial tool for addressing these issues at early career stages (e.g., Russell \& Dye 2014). 
However, the pandemic threatens to roll back the progress made on many other fronts by introducing a new set of barriers that may limit the participation of already marginalized populations within academia. 
Recent surveys reveal that 1 in 10 young adults have had to relocate due to COVID-19 (Cohn 2021), with a growing number citing financial problems as the most important reason for the move. 
Furthermore, those working from home report having difficulty feeling motivated, meeting deadlines, and getting work done without interruptions, especially workers in education and scientific fields (Minkin 2021). 
There is also the issue of technological barriers, such as limited access to the resources needed to reliably work from home. 
These collateral consequences of the COVID-19 pandemic disproportionately affect people of color, women, people with low income, and intersections therein (Getachew et al. 2020; Parker, Minkin \& Bennett 2020; Barroso 2021). 
Students may find themselves excluded from or unable to accept opportunities due to location and pandemic restrictions, and unable to fully engage in academic and research activities due to technological or logistical barriers. 
These inequities will further drive the gap between marginalized and privileged academics well into the future unless steps are taken today to mitigate the damage done by the pandemic within the academic sphere.  

Founded in 2013 by the National Radio Astronomy Observatory (NRAO) and Associated Universities Inc. (AUI) in partnership with the National Society of Black Physicists (NSBP) and several universities across the United States, the National Astronomy Consortium (NAC) was devised as a means of addressing such inequities prior to the onset of the pandemic. 
The mission of the NAC is to increase the numbers of students from underrepresented and underserved groups in STEM through a system based on the \emph{Posse Foundation} model (Bial 2016, Grotheer 2019). 
Under normal circumstances, NAC students are chosen from a pool of applicants and divided into cohorts hosted by partner institutions. 
NAC students are given hands-on research experience through paid summer internships, during which they participate in weekly cohort building and professional development activities. 
At the end of the program, students have an opportunity to present their work to fellow NAC students and alumni across all partnered sites at the annual fall NAC conference. The NAC also provides long-term financial support (e.g., travel funding to professional conferences and graduate application fee waivers) and mentoring. 
One of the most important aspects of the NAC program is continued student  engagement with the program as NAC alumni and as peer- and near-peer mentors to future cohorts of NAC students. 
Thus, our students belong to an expansive network of support with mentors who are committed to helping NAC students navigate academia long-term. 

Typically, these opportunities are operated in-person, with students traveling to 1 of 7 current partner sites to conduct their research. 
These sites were encouraged to convert to an online format as the pandemic spread, which introduced obstacles similar to those now faced by universities and research institutions nationwide. 
In this paper, we address the challenges the NAC faced in transitioning to a fully virtual format for both the summer research portion and the annual NAC conference. 
The authors are graduate student alumni of the NAC who hold leadership roles as peer- and near-peer mentors within the program. 
Each was actively involved in the planning and implementing of NAC projects over the past year, including the 2020 NAC summer program and the annual conference. 
Here we detail our methods, evaluate the success of these efforts, and provide recommendations for future virtual programming.

\section{Survey Introduction}
\label{section:survey}

To provide context for the changes the NAC made to its programs and their outcomes, we report the results of the NAC Alumni Survey, which was adapted from the ``Staying in Science" Alumni Survey (Podkul et al. 2018) and performed from September through December 2020 via Qualtrics. 
The purpose of the survey was to better understand the experiences of NAC students and alumni in college and in research, especially within the context of the NAC and the services offered through the organization. 
The survey also captured, in part, the effect the COVID-19 pandemic had on participants' education and research situations. 
Participants received compensation in the form of one \$25 Visa gift card each for filling out the survey, which took approximately 20-30 minutes to complete. 
The survey was sent to all 72 NAC alumni and received 47 responses, 44 of which were completed in their entirety. 
Of the respondents, approximately 48\% (or 22 alumni) had graduated college, while 52\% were undergraduate students at the time the survey was conducted. 
Approximately 64\% of those who had already graduated (14 alumni) were enrolled in a graduate program, with all but one enrolled in a doctoral program. 
The remaining 24 undergraduate students all anticipated graduating by 2024, with one-third having planned to graduate in 2020 at the time of the survey. 

By design, NAC serves a diverse student population, as reflected by the survey results. 
Of the 41 respondents who answered the race and ethnicity portion of the survey, approximately 38\% identify as Black, one-third identify as white, and approximately 15\% selected ``Other." The least represented groups were East Asian (6\%), Native American and/or Alaska Native (4\%), South Asian (2\%), and Native Hawaiian and/or Pacific Islander (2\%). 
Nearly 32\% identified as Hispanic or Latinx. 75\% out of 40 respondents reported English being the only language spoken with their immediate family (compared to 15\% stating that English is never spoken with family). 44\% of 41 respondents have at least one parent who was born outside of the United States. 

It should be reiterated, as previously explained, that minoritized students are more likely to be affected long-term by the pandemic (e.g., Getachew et al. 2020, CDC 2021). 
Ultimately, these students and alumni make up a small sample size, so the results of this report and its applicability to a wider community should be approached with caution. 
We note, however, that every respondent reported that they would recommend the NAC to a friend, and more than 95\% stated that they felt that the NAC was useful. 
Considering this, and taking the wide diversity of the respondents into account, we believe the results of this survey and the outcomes of our remote summer research program and conference during the pandemic can serve as a useful example for other programs looking to improve and provide support to marginalized members of their community.

\section{Summer Research Experience}
\label{section:reu}
In the early months of 2020, NAC program leaders, like others around the U.S., were busy reviewing applications, solidifying projects and logistics, and making offers to students without knowing the impending, drastic impact of the novel coronavirus. By March, it became clear that we would have to significantly adjust our program model to ensure the safety of our students, mentors, and the general public. Of the students who responded to the NAC Alumni Survey, 15 out of 44 respondents reported that their ability to participate in summer research had changed due to the COVID-19 pandemic. 3 respondents had a summer internship/job cancelled, and 14 had a summer internship/job moved online. Because cancellations could be detrimental in the long-term, especially to marginalized students, it was of the utmost importance that we find a way to adjust the program, which was already set to begin in June and last from 8 to 12 weeks, to accommodate the circumstances rather than eliminating the summer program altogether. Here, we break down the changes we made to the summer research program into two categories: 1) research-related changes pertaining to the equipment and skill-building opportunities provided for the students and the logistics of the program as a whole; and 2) the addition of or changes to personal accommodations such as cohort-building exercises and wellness checks aimed at addressing the physical, mental, and emotional strains of working through the pandemic and the tense political climate of 2020.

\subsection{Research-Related Changes}
\label{sub:research-changes}

With lockdowns beginning in March in the U.S., the NAC program had short notice to make the necessary safety changes. One major challenge was redesigning the program to accommodate the living situations of the students as the world moved into quarantine. Typically, students spend the summer in the city of their research site, which gives them the opportunity to connect with their cohorts and experience the local area. This year, none of the NAC partner sites were able to accept students in person. Because of international travel restrictions, one student in particular was not able to get back to the U.S. for the summer, and the international time zone difference ultimately affected their ability to participate in the summer research experience with the NAC. In addition, nearly half of the alumni (48\% out of 44) reported a change in their housing situation in 2020 due to the COVID-19 pandemic. 34\% lived at home with family, while nearly 56\% lived on or near their university campus during the 2020-2021 academic year. Prior to the start of the summer, each student was asked what their individual living situation would be for the duration of the research program. Understanding each student's circumstances ensured we could address logistical and technological needs beforehand. Each student-mentor pair developed a communication and research plan independently, so that each student's unique schedule could be worked around as needed.

According to the Alumni Survey, 9 out of 39 students reported that they did not have access to the technologies they needed to study or do school-related work during the 2020 semesters. In preparation for the shift to all-remote, the NAC communicated with each student to ensure they had the technological resources required to participate in the all-virtual platform. We installed necessary software and shipped laptops and electronic accessories to students’ summer residences. We also hosted virtual Python coding boot camps every other week to help our students learn the skills necessary to do their research. The NAC Alumni Survey conducted in the fall reflects the importance of these programs: the ability to use computer software such as MATLAB or Python and to analyze data were ranked the most helpful skills students learned in preparation for graduate school and other research experiences.

An important feature of the NAC summer research program is the opportunity to engage in weekly professional development meetings and exercises. In the past, each site’s students would meet in-person weekly, and then the full cohort from all sites would meet virtually a few times throughout the summer. During the pandemic we were able to continue providing these opportunities virtually and extend them to the entire cohort weekly. With the loss of in-person connections, having the full cohort meet weekly enhanced the sense of community by providing shared experiences. This year, our events included creating CVs and personal statements, a career/graduate student panel, and talks about science communication, social media presence, and imposter syndrome and how to tackle it.

\subsection{Personal Accommodations}
\label{sub:personal-changes}

Our program recognizes that this is a stressful time that we are asking students to work through. Vindegaard \& Eriksen Benros (2020) found that patients with preexisting psychiatric disorders reported worsening psychiatric symptoms as a result of the pandemic, and a decrease in psychological well-being was observed in the general public overall. These pandemic-related stressors were compounded by other sources of trauma that arose throughout 2020, including the murder of George Floyd and the ensuing Black Lives Matter protests for racial justice, and the tense political atmosphere of the 2020 U.S. elections. It is vital to address the students’ mental health in order to ensure students had the support they needed to thrive both personally and academically. We therefore enlisted a mental health provider to hold sessions on stress reduction and coping strategies and to be a resource for overwhelmed students. Many students attended the weekly session as well as scheduled individual appointments with the counselor. This initiative proved to be helpful, and we plan to continue offering access to a mental health provider even after we move away from the virtual setting. 

The NAC also recognizes the importance of the interpersonal support systems students build with others in their cohort. In previous years, individual cohorts for all seven sites met separately for weekly check-ins and professional development. Moving to an all-virtual platform gave us the opportunity to bring all the cohorts together for weekly check in-meetings to help build the sense of camaraderie and community, even as many students were confined to isolation. The students were also offered fun activities to take their minds off work, such as murder mystery nights, an escape room, dinners via Zoom, and a riddle night. Later, during the semester, we also held study sessions and occasional catch-up sessions to continue cohort interaction. The NAC Alumni Survey reflects the impact of these efforts: every student response noted that support from peers made the respondent feel more comfortable in STEM, despite the fact that 35\% (out of 43 respondents) reported rarely encountering career scientists of a similar race or ethnicity, and more than one-quarter report only occasionally (or less often) encountering career scientists of the same gender.

\section{NAC Conference}
\label{section:conference}

As the pandemic dragged on, the annual NAC conference also needed to shift to a virtual format to maintain safety and comply with travel restrictions. 
However, unlike with the summer research experience, the NAC had several months to prepare. 
Given the new format, the NAC wanted to take advantage of opportunities that would not be possible during our traditional in-person conference. 
A virtual conference format would lose in-person benefits such as formal networking, casual interactions, and the dedicated time allocation. 
However, new opportunities such as the ability to invite distant speakers and reallocate savings from travel and accommodation costs to other endeavours were now possible. 
The NAC focused its restructuring of the conference around the students’ new needs and challenges during the pandemic. 
All alumni were given an opportunity to design the conference or put forth suggestions for this reimagined virtual conference. 
Ultimately, nine alumni volunteered to be on the Organizing Committee (OC) and three of those volunteered to serve as the committee Co-chairs. 
The authors of this paper were all members of the OC, and one was a Co-chair.  
The NAC offered a modest stipend in exchange for their time and efforts, and to ease student financial burdens resulting from the pandemic. 

NAC alumni that joined the conference as organizers or participants received both direct and indirect benefits from the first virtual conference. 
OC students improved their leadership and organizational skills through exercising authority over the conference content and structure. 
By giving students a voice in the decision making process, the NAC provided a powerful platform for them to enact their ideas and foster student representation. 
Post-conference, OC members all expressed that the experience gave them valuable insight on event planning and allowed them to improve their management, teamwork, and other soft skills. 
Alumni participation increased for the virtual conference, as we had 16 alumni registrants compared to the 2019 in-person conference with 9 alumni registrants. 
This resulted in a larger audience for general talks, more program content, greater networking opportunities, and strengthened the community of our program. 
No matter their level of involvement, students who participated in the virtual conference garnered new knowledge and skills along with professional accomplishments applicable to their curriculum vitaes. 

Communication for a virtual conference is critical, so we created a centralized website using Cvent software. 
Through the website, the organizers could update the schedule of events, introduce themselves and speakers, provide general information, and see expected participation levels for each session. 
For participants, the website served as a resource with information about events, technology platforms utilized, channels of communication for help, and registration for individual sessions. 
There were no registration fees, and the only deadline was to ensure conference supplies for specific sessions could be shipped on time. 
New registrants were admitted beyond this deadline; however, they would be responsible for supplying their own materials as needed. 
The development of this website and shipment of supplies created challenges that we further address in \S\ref{sub:conference}.

The goals of the virtual conference remained the same as the in-person version: to give students the opportunity to present research conducted during the summer program, engage with the community, and network. 
Throughout August, the OC met weekly and planned a schedule of programming that balanced the three main conference goals with supporting students through the struggles they were facing during the on-going pandemic. 
These issues included differences in time zones, the encroachment of individual daily responsibilities, and the emotional and physical effects of ``Zoom fatigue" (Lee 2020). 
The conference was divided into two-hour long evening sessions on Thursdays and Sundays for five weeks to approximate the 20 hours we would have normally spent together for in-person content. 
This format ensured weekly attendance would not be overly burdensome to participants, as it was limited to a maximum of 4 hours weekly. 
However, choosing presentation times that accommodated various time zones was challenging and is a point we address later in \S\ref{sub:conference}. 

The structure of each two-hour session was divided equally into two sections: an open general session for research talks and networking events, followed by a private session for the students with guided cohort-building activities. 
This allowed us to accommodate the schedules of institutional partners by limiting demands on their weekly time while maintaining student engagement in a smaller and less formal setting. 
Students within the NAC program were expected to attend both sections. 
We used one virtual platform, Zoom, for both the public and student portions to lessen the technological demands on participants. 
For consistency, one Zoom link was used for all of the general sessions, while all cohort building sessions used another Zoom link throughout the conference. 
We also maintained consistent start times across the weeks to encourage timely attendance of participants. 

The general sessions were open to all registrants and focused on science talks, professional development, and formal networking. 
These events mirrored the typical schedule of our in-person conference. 
We had fifteen students give talks describing their summer research, two alumni research talks, four alumni presenting professional development workshops, five invited speakers, welcome and closing sessions by the organizing committee, formal networking events, and a career night with various institutional representatives discussing their career paths and potential opportunities for our students. 
The general sessions were recorded for asynchronous access. 
Because a virtual conference meant speaker slots could be spread over a wider time frame, there was greater flexibility for scheduling invited speakers than an in-person conference would provide. 
We believe this increased our opportunities for securing desired speakers. 
However, the pandemic may have also temporarily increased many speakers' availability, so it is difficult to discern if this is a true advantage a virtual conference would hold over in-person conferences once the pandemic has ended. 
Even so, a virtual conference inherently removes travel time and costs, so speakers have the potential to participate in multiple events at different locations, and organizers can afford to invite distant speakers who would normally not be able to attend do to their budget constraints. 

Participants who registered as students or alumni had access to private sessions geared towards cohort building, which were particularly emphasized this year due to the remote format. 
These sessions included seasonal activities such as pumpkin painting, structured games like murder mystery and charades, and casual conversations while coloring or visiting interesting websites. 
Certain activities required supplies to be sent to attendees’ locations prior to the event, which posed a logistical challenge discussed further in \S\ref{sub:conference}. 
These sessions, which provided shared experiences amongst the students, helped to forge relationships. 
Private sessions also gave us the opportunity to address the ongoing struggles of the pandemic and the tense social and political climate in the US. 
We held appropriately timed counseling sessions, and every meeting was a safe space to discuss school, life, and pandemic stress among peers. 
Due to the nature of the conversations and informal setting we wished to cultivate, cohort building sessions were not recorded, so attendance was crucial. 
To alleviate some of the time demand, incentivize participation, and address pandemic-related food insecurity (Laska 2020), we also offered `Sunday dinner' meal gift cards to students registered for Sunday events. 
Our registered student participation and engagement remained consistent throughout the conference, but our Zoom attendance records show that there was notably higher participation among students for Sunday sessions. 

\section{Challenges \& Recommendations}
\label{section:challenges}

The COVID-19 global pandemic posed new challenges and amplified existing ones. In the following subsections we identify these challenges, evaluate our responses, and offer recommendations to improve results. We summarize these recommendations in \S\ref{section:conclusions}. The brief timeline between the declaration of the pandemic and the start of our summer program meant we faced many urgent logistical challenges, and our initial focus was to facilitate participation. Once the ability to participate was managed, we shifted to address the amplified emotional, physical, and financial needs of our students. This shift in focus carried into our annual NAC conference as the academic community grew accustomed to virtual formats. Unlike the summer research experience, we had sufficient advance notice to contemplate the different logistics of a virtual conference. Overall, for both summer and fall programming the methods of communication and the differences in participants’ physical locations were the most important factors for consideration. Transitioning from in-person to fully-remote translates to a shift in focus from centering programming to centering the participants’ needs. We can no longer expect participants to adjust to our programming schedule; rather, we must build attractive programming that can also conform to participants’ lives.

\subsection{Summer Research Experience}
\label{sub:reu}

The move to an all-virtual format for summer research posed several obstacles, including technology issues (lack of computers or reliable internet), new work environments (which were not always optimal), difficulty working with mentors from afar via Zoom, and the struggle to build cohort communities. To rectify these issues, we discuss changes implemented to move to a virtual summer research experience and recommendations for specific issues.  \smallskip

\emph{Student Housing}: Due to travel restrictions, students were not able to travel to their cohort site, nor were they able to room with cohort students as they typically would have. This poses two main issues. First and foremost, the sudden switch to a virtual summer research program introduced a lack of housing security for the summer. With the program moving online, some students had to re-evaluate their summer plans and find new housing options given their unique circumstances. Second, because students did not see each other face-to-face or even outside our weekly meetings, friendships were more difficult to form. Informal after-hours interactions are vital to building a strong, supportive community amongst the students, which is a core tenet of the NAC model. Our recommendation is to continue to bring students together to a cohort site and to house students together as soon as it is safe. This allows students to experience a new city, meet students from the program, and guarantee housing for the summer. \smallskip 

\emph{Student Finances}: Financial security is a well known issue for many students. Typical summer research programs run from 8-12 weeks. For some, there are periods of time when students will not be earning an income due to a late start and/or early finish to their program. Because of this, students have to stretch their summer stipend to make up for the difference. Additionally, students are sometimes expected to front the money for travel to their site before they are even paid their first stipend. To address these financial issues, which often have a greater effect on marginalized communities (Parker et al. 2020, Yaphet et al. 2020), we offer two possible suggestions. First, offer to extend the internship for the entire duration of the summer for students who desire the extra financial support and research time. Second, we suggest making it a high priority to have travel expenses paid directly from the summer research institution (as is standard NAC practice) rather than having students organize their travel and pay the money upfront. \smallskip

\emph{Mentorship}: While we know support and mentoring from advisors and peers is critical, the pandemic reiterated its importance. A multi-faceted support system should be implemented for post-pandemic summers, but especially for a virtual summer. One change the NAC plans to implement next year to build relationships with both primary and community mentors is to introduce informal “science tea” gatherings with small presentations given by mentors. These presentations could be about their research, their career paths, or other topics. Informal sessions like these help to de-emphasize the hierarchy of the advisor-student relationship. Furthermore, building relationships with a wider community of career researchers opens doors for students to increased opportunities for mentorship. We recommend the addition of  “intermediate mentors'' who are more senior students or junior scientists and whom summer students may feel more comfortable coming to for help or advice. This fosters peer-mentoring and community support, which is an important aspect contributing to a student’s growth in the field. For the NAC, this would include previous alumni of the program. For other programs that don’t have alumni as a resource, such as programs run by universities or observatories, the host institution could offer modest stipends to graduate students or junior scientists to participate.  \smallskip

\emph{Engagement}: Because of the challenging times the students faced this summer, we had to work more deliberately to keep the students engaged. We present two recommendations here for helping address engagement. First, we recommend a pre- and post-summer survey to gauge what students want, expect, and the overall effectiveness of the program. Mid-summer evaluations are also helpful because they allow for more immediate adjustments to better suit the students’ needs as they arise. Additionally, we recommend keeping track of attendance, to 1) ensure that students are doing fine overall and 2) gauge participation and interest in specific topics amongst the group, which may be used to plan for future sessions. \smallskip

\emph{Communication}: The majority of communication between students, mentors, and administrators was via email during this summer. While a Slack channel was set up for students to ask questions and relay information, it was not utilized very heavily. We recommend choosing one platform for meetings (i.e. Zoom, Google Meets, Skype) and one platform for official communication (i.e. Slack, Discord). We also supported students incorporating social media and other platforms for informal communication and science outreach. We brought in speakers to address Education and Public Outreach (EPO) in a virtual setting and both personal and professional social media presence. With much of our lives being confined to the internet, students are looking for new ways to stay involved, and the use of social media is important to this involvement.  \smallskip

\emph{Emotional Support}: This past summer, we emphasized the importance of addressing imposter syndrome directly and frequently. Many of us struggle with imposter syndrome, including a reported 93\% of NAC students based on survey results, but it is often considered taboo to discuss in social settings (Gardner et al. 2019, Jaremka et al. 2020). Imposter syndrome is a significant topic in academic settings, and addressing it early in a student’s career will help them as a student, scientist, and person. We advise at least one formal discussion about imposter syndrome, preferably led by an invited specialist, as well as multiple informal discussions regarding personal struggles amongst the students and mentors.  Furthermore, students may benefit from access to one-on-one counselling (which we provided this year and will continue to provide in the future). We recommend that institutions provide students with similar access, if possible. 

\subsection{NAC Conference}
\label{sub:conference}

Overall, the NAC program’s first transition to a fully-virtual conference was challenging but mostly successful. The obstacles this shift introduced can be broken into three main categories: scheduling, event supplies planning, and website creation. Here we discuss the choices we made and the changes we would recommend for a smoother virtual event. \smallskip

\emph{Scheduling}: A significant concern while planning the virtual conference was setting a reasonable and consistent event schedule that took into consideration both the time zones of all participants, which included locations as distant as India and Hawai'i, and other daily obligations of the participants. We decided to host activities from 6 to 8 PM Eastern Standard Time over the course of five weeks, since most attendees were located across the mainland US. To address the farther time zones, we instituted programming on weekends as well. Our approach of static meeting days and start times throughout the weeks was intended to provide participants with a consistent meeting pattern to avoid absences. However, this resulted in consistently inconvenient overnight or early morning hours for our farthest participants. Recorded sessions ensured the general talks were available asynchronously, but there was no recorded option for cohort-building activities. In the future, to ensure events are accessible to all participants, we suggest alternating meeting times between mornings and afternoons or offering the same events across multiple days if possible. In addition, the five week duration of the conference became minorly taxing. Though attendance remained consistent throughout the consecutive weeks, there was an emotional toll commonly expressed in the later weeks. For future virtual conferences, we recommend limiting the overall duration to a maximum of three weeks. \smallskip

\emph{Event Supplies Planning}: Another concern was ensuring that all materials required for cohort-building events were shipped to students and alumni in time.  While there was no registration deadline for the conference as a whole, we instituted a deadline for these events so that we could have an accurate account of how many packages were needed and where they would be shipped. Events requiring special supplies were scheduled toward the end of the five week conference to allow ample shipping time. For future events, we recommend that organizers determine if supplies would be helpful to their program goals as early as possible and prioritize the planning of those events first.  \smallskip

\emph{Website Creation}: The creation of a conference website was challenging to the organizing committee due to a lack of experience with website development software. The content of the website, including the master schedule and session descriptions, had to be generated by the committee prior to the opening of registration. A subcommittee was tasked with learning the software and testing user interactions to ensure smooth registration once the site was live. The additional work of learning new software and writing content descriptions was initially underappreciated, so we recommend this is begun at the earliest stage of planning possible. \smallskip

While there are many challenges, a benefit to the remote format of both our summer research experience and the virtual conference is that it introduced the possibility of engagement by students and others who may otherwise be left out of summer opportunities due to personal, financial, or medical reasons. Virtual platforms allow people to work from anywhere, simultaneously or asynchronously. While our recommendations may alleviate some challenges that pose a barrier to engagement by marginalized peoples, unique situations will still arise. We recommend that programs keep a virtual option open, even beyond the pandemic when in-person functions can resume. This may include maintaining a license for virtual platforms (e.g. Zoom) and introducing infrastructure for recording and archiving in-person events. Doing so will broaden the accessibility of any program in the long-term.

\section{Conclusions}
\label{section:conclusions}

After over a year of working through a pandemic, the U.S. (and the world) is still navigating through unknown territory. While these recommendations come from reflecting on the past summer and brainstorming to improve future summers, it is not to say that all recommendations are entirely feasible for every program. With the next round of summer research programs in their early planning stages, programs are now more equipped to adapt their programs to ensure the same valuable experience for students virtually. In summary, moving forward we recommend the following:

\begin{itemize}
    \item Do not cancel summer research experiences. Virtual summer research programs are better than none, though it is preferred that programs send students to sites and house students together, where possible and when safe.
    \item Identify all technological requirements (e.g. virtual platform, website, etc.) up-front and prioritize setting these up first. Make sure all participants have access to these resources far in advance. 
    \item Minimize the number of digital platforms/programs/applications used across all functions. A single platform for meetings and a single platform for official communications are sufficient in most cases.
    \item Identify all physical materials required for events and ship them as soon as possible. Additionally, make a list of materials and the dates they will be used to provide to participants who miss the deadline to have materials shipped to them.
    \item For in-person research opportunities, provide travel funding to (and from, if short-term) research experiences up-front, or secure travel arrangements for students directly.
    \item For summer opportunities, offer extended internship dates or `bridge stipends' that account for more of the summer to alleviate the financial burdens of the most vulnerable students served by the program. 
    \item Provide informal networking opportunities throughout the summer, such as ``science tea" sessions.
    \item Institute an ``intermediate" or secondary mentorship system, preferably including graduate students or early career researchers that students may feel comfortable turning to for support.
    \item Provide frequent opportunities for feedback before, during, and after internship programs so that changes may be made to center the needs of the participants.
    \item For partially or fully virtual conferences, schedule programming with staggered start-times across multiple days to accommodate time zones and busy schedules. Record non-sensitive sessions and offer repeat events.
    \item If scheduling a conference over a long time frame, limit the total duration to a maximum of around 3 weeks.
    \item Emphasize the importance of emotional support in addition to professional support. Discuss the prevalence of impostor syndrome often, and provide students (and all members of the community) with additional avenues for emotional support (i.e. access to counseling). Always practice compassion with students who have fallen behind for whatever reason. 
    \item Maintain a virtual option beyond the end of the pandemic to make programs and conferences more accessible indefinitely.

\end{itemize}

With these recommendations and the individual adaptations of each entity, institutions may offer more engaging content to their students, provide greater accessibility to their programs, and allow students to gain those invaluable resources and opportunities vital for their future endeavors.

\section*{Acknowledgments}

The authors wish to thank Lyndele von Schill, Karen Prairie, Isabelle Marsh, Jessica Burns, and NAC administration for their unwavering support throughout the years, and to thank NAC alumni for their participation in and dedication to the development of a long-term supportive community. NAC Alumni Survey results were collected and analyzed by Meredith Graham, and we are appreciative of all survey participants. For their work developing and facilitating the annual fall conference, we also wish to thank our fellow Organizing Committee members Debora Mroczek, Maryam Hami, Sinclaire Manning, Tiffany Christian and additional Co-chairs Alia Wofford and Antonio Porras. 

The NAC is made possible through the generous support of the National Radio Astronomy Observatory (NRAO) and the National Science Foundation (NSF). The National Radio Astronomy Observatory is a facility of the National Science Foundation operated under cooperative agreement by Associated Universities, Inc. This material is based upon work supported by the NSF under Cooperative Agreements No. 1647375 and 1647378. Our thanks are extended for our ongoing partnership with the National Society of Black Physicists (NSBP). We would also like to acknowledge our current partner sites’ funding of and mentorship during the NAC summer research program. Summer research partner sites include: Michigan State University, NRAO Central Development Laboratory (CDL), NRAO Headquarters, NRAO Domenici Science Operations Center (DSOC), Princeton University, Space Telescope Science Institute (STScI), \& University of Wisconsin-Madison. Any opinions, findings, and conclusions or recommendations expressed in this material are those of the author(s) and do not necessarily reflect the views of the National Science Foundation.



\newpage
\begin{references}


Barrosa, Amanda. (2021). For American couples, gender gaps in sharing household responsibilities persist amid pandemic, Pew Research Center. Retrieved from 
\href{https://www.pewresearch.org}{https://www.pewresearch.org}\\

Bial, Deborah. (2016). Meritocracy Stalled: Diversity, Higher Education, and America's Leadership. Retrieved from \href{https://www.possefoundation.org}{https://www.possefoundation.org}\\

Brookshire, B. \& Raloff, J. (2020). Explainer: What is a Mentor? Science News for Students, Science and Society. Retrieved from\\ \href{http://www.sciencenewsforstudents.org}{http://www.sciencenewsforstudents.org}\\

Centers for Disease Control and Prevention (CDC). (2021). Hospitalization and Death by Race/Ethnicity. Retrieved from \href{https://www.cdc.gov/}{https://www.cdc.gov/}\\

Chen, X., \& Soldner, M. (2013). STEM Attrition: College Students' Parths into and out of STEM Fields. Statistics Analysis Report. U.S. Department of Education, National Center for Education Statistics. Retrieved from \href{https://eric.ed.gov/}{https://eric.ed.gov/}\\

Cohn, D'vera. (2021). As the pandemic persisted, financial pressures became a bigger factor in why Americans decided to move, Pew Research Center. Retrieved from \href{https://www.pewresearch.org/}{https://www.pewresearch.org/}\\

Gardner, https://eric.ed.gov/R. G., Bednar, J. S., Stewart, B. W., Oldroyd, J. B., \& Moore, J. (2019). “I must have slipped through the cracks somehow”: An examination of coping with perceived impostorism and the role of social support. Journal of Vocational Behavior, 103337. \href{https://www.sciencedirect.com/science/article/pii/S0001879119301095}{doi:10.1016/j.jvb.2019.103337}\\

Getachew, Y., Zephyrin, L., Abrams, M. K., et al. (2020). Beyond the Case Count: The Wide-Ranging Disparities of COVID-19 in the United States, Commonwealth Fund. \href{https://www.commonwealthfund.org/publications/2020/sep/beyond-case-count-disparities-covid-19-united-states}{doi:10.26099/gjcn-1z31}\\

Gonzalez T, de la Rubia MA, Hincz KP, Comas-Lopez M, Subirats L, Fort S, et al. (2020) Influence of COVID-19 confinement on students’ performance in higher education. PLoS ONE 15(10): e0239490. \\
\href{https://journals.plos.org/plosone/article?id=10.1371/journal.pone.0239490}{https://doi.org/10.1371/journal.pone.0239490}\\

Grotheer, Emily. (2019). The Posse Foundation: Annual Report 2019. Retrieved from \href{https://www.possefoundation.org/}{https://www.possefoundation.org/}\\

Jaremka LM, Ackerman JM, Gawronski B, et al. Common Academic Experiences No One Talks About: Repeated Rejection, Impostor Syndrome, and Burnout. Perspectives on Psychological Science. 2020;15(3):519-543. \href{https://journals.sagepub.com/doi/abs/10.1177/1745691619898848?journalCode=ppsa}{doi:10.1177/1745691619898848}\\

Laska, M. N., Fleischhacker, S., Petsoulis, C., Bruening, M., \& Stebleton, M. J. (2020). Addressing College Food Insecurity: An Assessment of Federal Legislation Before and During Coronavirus Disease-2019. Journal of nutrition education and behavior, 52(10), 982–987.\\ \href{https://www.ncbi.nlm.nih.gov/pmc/articles/PMC7450237/}{https://doi.org/10.1016/j.jneb.2020.07.001} \\

Lee, Jena. (2020). A Neuropyschological Exploration of Zoom Fatigue. Psychiatric Times. Retrieved from \href{https://www.psychiatrictimes.com/}{https://www.psychiatrictimes.com/} \\

Minkin, Rachel. (2021). Even in industries where majorities can telework, some face challenges working from home during pandemic, Pew Research Center. Retrieved from \href{https://www.pewresearch.org/}{https://www.pewresearch.org/}\\

Parker, K., Minkin, R., and Bennett, J. (2020). Economic fallout from COVID-19 Continued to Hit Lower-Income Americans the Hardest, Pew Research Center. Retrieved from \href{https://www.pewresearch.org/}{https://www.pewresearch.org/}\\

Podkul, T., Gupta, P., Chaffee, R., and Hammerness, K. (March 12, 2018). Staying in science: An examination of youth pathways using social network theory and analysis. National Association of Research in Science Teaching Annual Conference. Atlanta, GA.\\

Russell, H. M., and Dye, H. A. (2014). Promoting REU participation from students in underrepresented groups, Involve, a Journal of Mathematics, Vol. 7, No. 3., \href{https://scholarship.richmond.edu/cgi/viewcontent.cgi?article=1156&=&context=mathcs-faculty-publications&=&sei-redir=1&referer=https%253A%252F%252Fwww.google.com%252Furl%253Fq%253Dhttps%253A%252F%252Fscholarship.richmond.edu%252Fcgi%252Fviewcontent.cgi%253Farticle%25253D1156%252526context%25253Dmathcs-faculty-publications%2526sa%253DD%2526source%253Deditors%2526ust%253D1617830353344000%2526usg%253DAOvVaw2NiIPu362TJEdASmu2OhyM#search=%22https%3A%2F%2Fscholarship.richmond.edu%2Fcgi%2Fviewcontent.cgi%3Farticle%3D1156%26context%3Dmathcs-faculty-publications%22}{doi:10.2140/involve.2014.7.403} Published on 28 April 2014.\\

Sloan, V., R. Haacker, R. Batchelor, and C. Garza (2020), How COVID-19 is affecting undergraduate research experiences, Eos, 101, \\
\href{https://eos.org/science-updates/how-covid-19-is-affecting-undergraduate-research-experiences}{https://doi.org/10.1029/2020EO145667} Published on 18 June 2020.\\

Vindegaard, N., \& Erikson Benros, M. (2020). COVID-19 pandemic and mental health consequences: Systematic review of the current evidence. Brain, Behavior, and Immunity, Vol. 89, 531. \href{https://www.sciencedirect.com/science/article/abs/pii/S0889159120309545?via%3Dihub}{doi:10.1016/j.bbi.2020.05.048}\\





\end{references}

\end{document}